# 3D Printed Architectured Silicones with Autonomic Self-healing and Creep-resistant Behavior

Stefano Menasce, Rafael Libanori, Fergal Coulter, André R. Studart

Complex Materials, Department of Materials, ETH Zürich, 8093 Zürich, Switzerland

**Abstract**

Self-healing silicones that are able to restore the functionalities and extend the lifetime of soft devices hold great potential in many applications. However, currently available silicones need to be triggered to self-heal or suffer from creep-induced irreversible deformation during use. Here, we design and print silicone objects that are programmed at the molecular and architecture levels to achieve self-healing at room temperature while simultaneously resisting creep. At the molecular scale, dioxaborolanes moieties are incorporated into silicones to synthesize self-healing vitrimers, whereas conventional covalent bonds are exploited to make creep-resistant elastomers. When combined into architectured printed parts at a coarser length scale, layered materials exhibit fast healing at room temperature without compromising the elastic recovery obtained from covalent polymer networks. A patient-specific vascular phantom is printed to demonstrate the potential of architectured silicones in creating damage-resilient functional devices using molecularly designed elastomer materials.

**Introduction**

Self-healing polymers are attractive materials because of their ability to recover structure and properties after physical damage, thereby enhancing the lifetime of functional devices. Several physical and chemical processes have been exploited to enable self-healing in polymers, including interchain diffusion of macromolecules, molecular rearrangement in covalent or supramolecular polymers, shape-memory effects, and triggered release of encapsulated reactive species. [1] Polymer networks crosslinked via reversible covalent or supramolecular bonds are particularly interesting due to the possibility to achieve intrinsic self-healing capabilities through molecular design. Supramolecular chemistry has been extensively used to design polymer networks with weak reversible bonds involving, for example, π-π stacking, cooperative hydrogen bonding, metal-ligand coordination, and guest-host interactions. [2] The design space has also been extended to stronger bonds by employing dynamic covalent moieties, such as disulfides, Diels-Alder adducts, imines and boronic esters. [3] In contrast to the weak reversible bonds typically present in supramolecular polymers, covalent adaptable networks containing strong and dynamic bonds allow polymer networks to change their topology through degenerate and dissociative exchange reactions. More recently, degenerate covalent adaptable networks have been explored to fabricate dynamic polymers displaying a combination of thermoplastic and thermoset materials by providing a constant number of covalent crosslinks during temperature-induced topological change [4, 5] The viscoelastic and self-healing properties of these dynamic polymers, also called vitrimers, are governed by the glass transition ($T_g$) and the topology-freezing transition ($T_v$) temperatures of the polymer chains. While $T_g$ reflects the temperature above which the macromolecular chains become mobile, $T_v$ represents the minimum temperature needed for the degenerate exchange reactions to take place within experimentally accessible timescales.

Silicone vitrimers are enticing materials in the context of self-healing polymers because their $T_g$ is much lower than room temperature, which results in viscoelastic properties that depend primarily on the rate of degenerate exchange reaction ($T_v$) rather than on chain mobility ($T_g$). To harness this potential, degenerate dynamic covalent bonds involving dioxaborolanes moieties have been exploited to synthesize a broad range of silicone vitrimers with tunable $T_v$, viscoelastic properties and self-healing response. [5] [6] When $T_v$ is designed to lie below room temperature, such silicone vitrimers experience pronounced creep and autonomically self-heal through metathesis of dioxaborolanes as degenerate exchange reaction. Alternatively, external chemical or physical stimuli can be applied to increase the temperature of the silicone above $T_v$ and thus achieve triggered self-healing. Despite the broad design space available, current silicone vitrimers are limited either by their low creep resistance (when $T > T_v$) or by the need of an external trigger, such as temperature, pressure or light, to self-heal (when $T < T_v$). Such trade-off between autonomic self-healing and creep resistance has thus far hindered the broader use of silicone vitrimers in many applications.

Reconciling trade-off properties through structural design is a hallmark of biological materials. [7] Living organisms create materials such as seashells, wood and tendons that combine antagonistic properties by building exquisite hierarchical structures. Such structures show molecular, microstructural and architectural features at increasing length scales in the form of chemical gradients, controlled fiber orientation, tunable porosity and multilayered designs. [8] A prominent example are seashells, which reconcile stiffness and fracture toughness in a single material through a hierarchical layered design. [9] The incorporation of some of these biological design principles in synthetic systems has led to the development of bioinspired materials with functionalities that are not accessible in conventional synthetic systems. [10] While self-healing in biological materials is often controlled by living cells, some of the hierarchical structural features found in nature may provide guidelines for the design of synthetic polymers that are able to simultaneously resist creep and self-heal without external triggers. The byssus threads that mussels use to attach themselves to rocks is an inspiring example of a biological material that achieves antagonistic elastic recovery and self-healing properties by utilizing metal coordination bonds at the molecular level and a core-shell architecture at a coarser length scale. [11]

Here, we design and investigate silicone vitrimers and elastomers that are combined in a multimaterial printing process to generate architectured silicones with both creep-resistant and autonomic self-healing response. The creep resistance and self-healing properties are designed individually at the molecular scale using specific functional monomers during the polymer synthesis. These antagonistic properties are then reconciled in a single object by designing a bespoke multimaterial layered architecture at a coarser length scale. To print the architectured silicones, we first synthesize a dynamic chain extender containing dioxaborolane moieties, which is later incorporated into the polysiloxane chains of prepolymers through thiol-ene photopolymerization. Next, we develop a chemically-compatible resin formulation to yield a stiff, strong and elastic silicone elastomer. Creep resistance and self-healing behaviour of single layers and bilayers of printed silicones are quantified and interpreted using established analytical models. Finally, the architectured silicone is implemented in a demonstrator to illustrate how its creep-resistance and autonomic self-healing properties can be utilized to restore the function of a surgical phantom model of a damaged aortic root.

**Results and discussion**

Silicones combining self-healing capabilities and creep resistance are created using layered architectures with controlled geometry and spatial distribution of dynamic and permanent covalent bonds (Figure 1). We demonstrate the concept taking as an example a complex-shaped surgical phantom of the aortic root (Figure 1a). The phantom should mimic the self-healing and elastic properties of the natural aortic tissue without undergoing creep deformation. To fulfill these conflicting tasks, the wall of the phantom model features a trilayer elastomer architecture comprising an inner grid-like cross-linked layer sandwiched between two outer self-healing layers (Figure 1b).

In this design, the self-healing effect is imparted by the outer crosslinked silicone layers containing an optimized fraction of dynamic covalent bonds (vitrimer), whereas the creep resistance arises from the inner crosslinked silicone layer containing only permanent covalent bonds (Figure 1b). The self-healing and creep-resistant properties of the individual silicone layers are programmed at the molecular level using dithiols containing dynamic and permanent covalent bonds to form reversible and irreversible networks, respectively (Figure 1c).

Self-healing at room temperature is achieved using dioxaborolane-containing molecules as reversible, dynamic monomers. The dynamic boronic ester groups are incorporated through thiol-ene photopolymerization into a polysiloxane-based prepolymer that is terminated with reactive thiol groups. The high mobility of the polysiloxane chains compared to the backbone of most organic polymers is expected to reduce the thermal barrier necessary for the metathesis reaction to occur. [6] This leads to an intrinsic self-healing material that deforms and creeps at room temperature in a few minutes without any external stimulus such as temperature, UV or pressure.

The creep-resistant elastomer layer relies on the use of a thiol-terminated polysiloxane prepolymer containing only permanent covalent bonds to form the silicone network. [12, 13] The presence of only permanent covalent bonds provides not only creep resistance, but also enhances the mechanical strength, elastic modulus and elastic recovery of the silicone network compared to the self-healing counterpart. By utilizing thiol- and vinyl-based monomers for both elastomer and dynamic polymers, we ensure chemical compatibility between the self-healing and the creep-resistant layer, which is an essential condition for the processing and interlayer adhesion in the final architectured material.

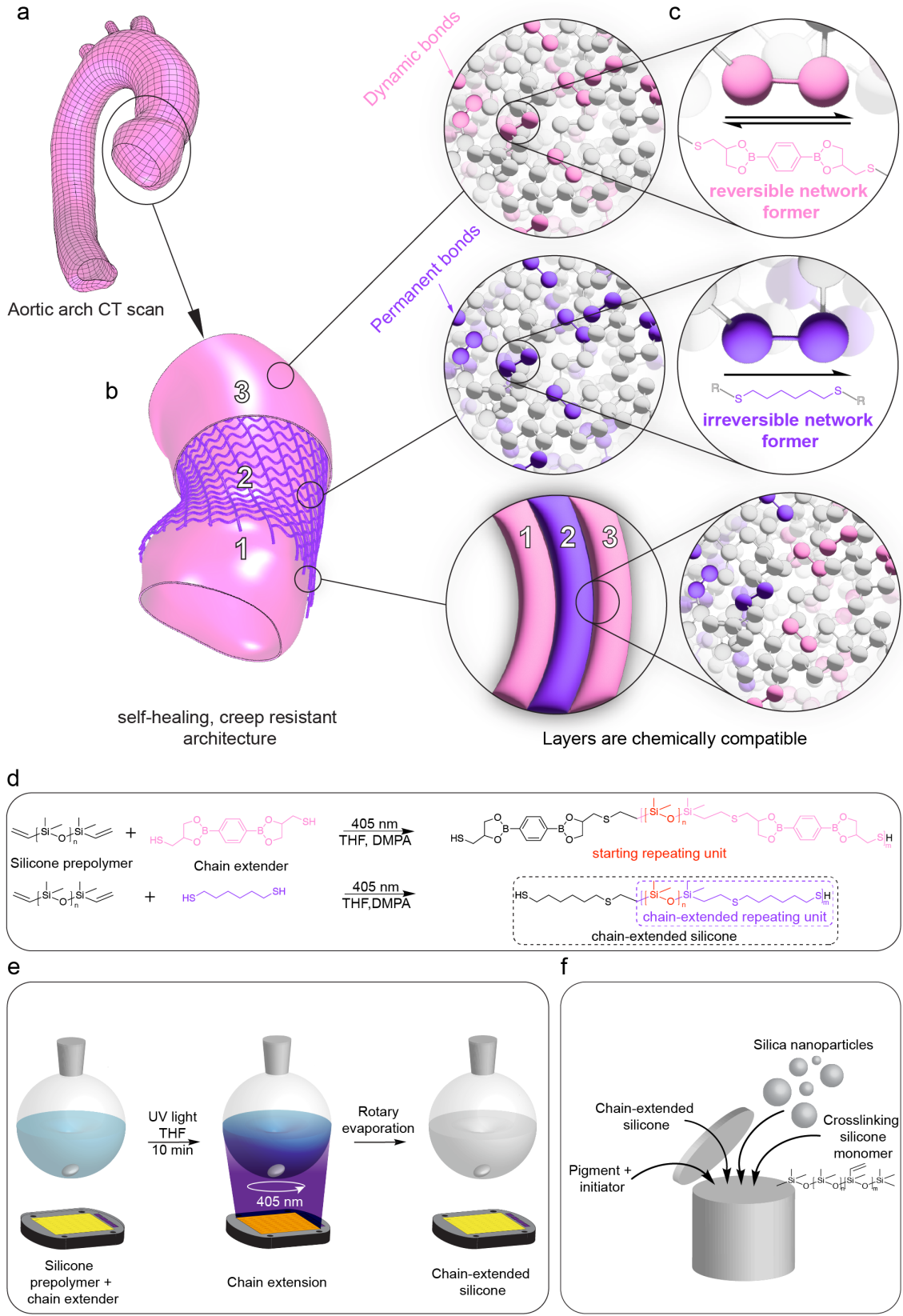

**Figure 1. Multiscale design of architectured silicones.** (a) At the architecture scale, a creep-resistant grid is sandwiched between two self-healing layers to generate the walls of a CT scan-derived aortic

arch. (b) At the molecular scale, dynamic and permanent covalent bonds are implemented in the silicone network to achieve the self-healing and creep-resistant properties of the individual layers, respectively. (c) Chemical compatibility between the layers is ensured by utilizing the same thiol-ene-based chemistry for the synthesis of the silicone network. (d) Chain extension reaction used to obtain the thiol-terminated dynamic (top) and permanent (bottom) prepolymers. (e) Experimental setup and protocol employed for the chain extension reaction. (f) Constituents used for the formulation of printable silicone inks.

Preliminary experiments indicated that the synthesis of thiol-terminated prepolymers is essential to build a strong and stiff three-dimensional silicone network. Indeed, direct mixing of short-chain dithiols with vinyl-terminated and polyfunctional vinyl-modified silicone prepolymers in the presence of photoinitiator does not form a strong and stiff percolating network upon photopolymerization. Because thiol-terminated long-chain silicones are challenging to synthesize, [12] we employed a pre-polymerization chain extension protocol to increase the molecular weight of the permanent and dynamic components that are used in the final elastomer formulation. [14]

The pre-polymerization reaction is performed in a separate step before the final polymerization of the silicone network (Figure 1d). In this step, the vinyl-terminated silicone prepolymer ($M_w$ = 9.4 kDa) is reacted with 1 to 2 equivalent excess of the thiol-based monomer in the presence of 2,2-dimethoxy-2-phenylacetophenone (DMPA) as photoinitiator [15] and tetrahydrofuran (THF) as solvent. Prepolymerization occurs via a thiol-ene click reaction triggered by UV light, which is known for its high yield and selectivity. [16] The reaction conditions were empirically optimized in terms of thiol:vinyl equivalent ratio, monomer concentrations and illumination conditions (Figure 1e).

Optimum thiol:vinyl ratios of 2:1 and 3:1 were used for the synthesis of the permanent and dynamic prepolymers, respectively. Unreacted, high-vapor-pressure short-chain thiols with permanent covalent bonds are removed by vacuum evaporation after the synthesis of the chain-extended permanent crosslinkers (Figure 1e). For the dynamic prepolymers, unreacted dioxaborolane-containing short-chain thiols were found to be crucial for to provide dynamic bonds with high exchange rates in the final self-healing network and were therefore maintained in the mixture after the pre-polymerization reaction.

The synthesis of thiol-terminated dynamic prepolymers was confirmed by the disappearance of $^1$H NMR peaks corresponding to the vinyl groups of the silicone-based prepolymer and the simultaneous appearance of $^1$H peaks corresponding to the protons next to sulfur atom in the sulfide (C-S-C) bond (Figure S1). Gel permeation chromatography revealed that the chain-extension reaction increased the molecular weight of the thiolated permanent silicone prepolymers by a factor of approximately 4 ($M_w$ = 30-36 kDa). Further experiments showed that molecular weights up to 70 kDa can be achieved using this chain-extension protocol (Figure S2).

Thiol-terminated silicone prepolymers are used to formulate photocurable and 3D-printable inks containing dynamic or permanent covalent bonds. In addition to the pre-polymerized components, the inks contain DMPA as initiator, pigments of different colors to distinguish formulations, a polyfunctional vinyl-modified silicone crosslinker as network former and hydrophobized fumed silica as reinforcing filler and rheology modifier (Figure 1f). Since these constituents are available in large quantities and the chain-extension reaction shows high yield, kilograms of inks with shelf-life of more than 3 months can be produced in the laboratory within about 24h.

Fumed silica contents of 25.9 and 37.5 wt% (35 and 60 **p**arts per **h**undred **r**ubber, phr) were used in inks containing dynamic and permanent covalent bonds, respectively. Such concentrations were found to be optimum to reach the rheological properties required for printing using the Direct Ink Writing (DIW) technique. Oscillatory rheology data show that the inks exhibit elastic properties at low amplitude strains, but can be fluidized above a critical yield shear stress (Figure 2a). Flow measurements at constant shear rate reveal yield stresses of 190 and 420 Pa for inks with dynamic and permanent covalent bonds, respectively (Figure 2b). These values lie above the threshold of 100 Pa typically needed to print distortion-free objects by DIW. [17] With a low-shear storage modulus ($G'$) higher than $10^4$ Pa, the inks are also sufficiently stiff to create grid-like structures that resist gravity-induced filament sagging[18]. Steady-shear and elastic recovery experiments were also performed to show that the inks display the shear-thinning behavior and gelation speed desired for DIW printing (Figure S3). This set of rheological properties allowed for the multimaterial printing of a broad range of silicone architectures with complex 3D geometries and multilayered designs.

To demonstrate the versatility and practical application of our ink, we printed a section of an aortic arch which could be used as a surgical training phantom (Figure 2c). Such a device would enable surgical students to practice various suturing and incision techniques in a more realistic scenario compared to state-of-the-art silicone phantoms. Because of the self-repairing nature of the silicone walls, students can repeat the task on the same device after a short period of time – even in an aqueous environment such as a mock circulatory system. To fabricate such device, the component was printed on a rotating mandrel with an irregular curved surface derived from a patient-specific scan, following a 3D printing process previously described [19]. The high yield stress of the inks enabled printing of a tessellated pattern of the permanently crosslinked ink over the surface of a previously deposited formulation with dynamic bonds (Figure 2d,e). Such a tessellated architecture has been developed to mimic the natural anisotropic compliance of an artery, where the vessel is more compliant in the circumferential than the longitudinal direction. [20]

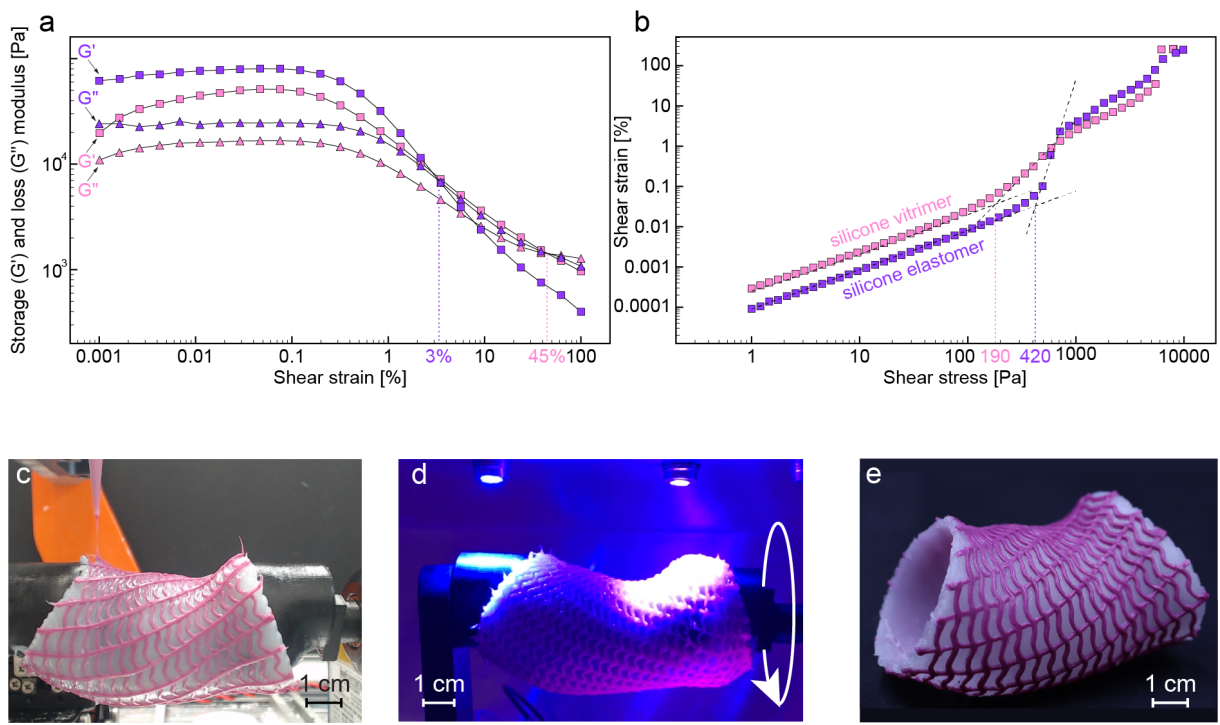

**Figure 2. Rheology of silicone inks and 3D printed objects.** (a) Flow curve and (b) viscoelastic properties of silicone inks with dynamic and permanent covalent bonds. (c) Example of architectured silicones printed by multimaterial direct ink writing. (d) Curing of the printed object under UV light. (e) Picture of the final printed and cured piece.

Formulations with permanent or dynamic covalent bonds were printed into dogbone samples to investigate the effect of the chemical bond nature on the creep resistance and self-healing properties of the silicones (Figure 3a and 3d). Creep tests were performed by applying a stress of 0.2 MPa for 90 mins to each specimen, following a standardized protocol. [21] Creep is quantified as the time-dependent displacement of the sample relative to its initial length upon tensile mechanical loading (Figure 3b). For the self-healing experiments, the specimens are cut in half and let at rest in close contact for a given amount of time before uniaxial tensile testing (Figure 3e). The strength of the self-healed sample is then compared to that of the pristine counterpart to calculate the self-healing efficiency.

The creep resistance and self-healing response of the printed specimens were found to be strongly affected by the presence of dynamic covalent bonds in the silicone network. In terms of creep, the silicone containing dynamic covalent bonds reaches displacement above 400% after 15 mins and shows continuously increasing deformation within the time window of the experiment (Figure 3c). In contrast, the displacement of the sample with only permanent covalent bonds remains lower than 10% irrespective of time. The evolution of the strain rate of these two materials during the test clearly indicates a constant creep for the network with dynamic covalent bonds, as opposed to a zero creep rate for the silicone with permanent crosslinks (Figure S4). Elastic recovery experiments under tensile mode reveal the predominantly elastic nature of the silicone with permanent bonds (Figure S5). Cyclic stress-strain data show that samples prepared with only permanent covalent bonds are able to recover up to 97% of their initial length, whereas the silicone with dynamic covalent bonds show large irreversible strain with an elastic recovery of only 20% after mechanical unloading.

The healing experiments indicate that dynamic covalent bonds are essential to recover the strength of cut samples, confirming the expected trade-off between creep resistance and self-healing capability (Figure 3f). The silicone with dynamic covalent bonds shows more than 60% strength recovery within only 5 minutes after being cut. The strength increases steadily and reaches its initial strength within 24 hours. Instead, the sample containing only permanent covalent bonds behaves as a conventional thermoset elastomer; once damaged, the two halves of the specimen do not adhere to each other and no self-healing occurs.

To gain more insights into the self-healing process of the silicone vitrimer, we compared our experimental data with the strength recovery expected assuming that healing occurs by the interpenetration of polymer chains across the two polymer surfaces. In this model, the polymer chains are assumed to diffuse via a double random walk process, leading to a theoretical strength of the interface ($\sigma$) that is given by: $\sigma = \sigma_0 + K n_0 (2D)^{1/4} t^{1/4}$, where $\sigma_0$ is the strength arising from wetting between the two surfaces, $K$ is a constant, $n_0$ is the number of constraints per unit volume of polymer, $D$ is the reptation diffusion coefficient and $t$ is the elapsed time. [22] The first term of this equation represents the wetting contribution to the recovered strength, whereas the second term corresponds to the strength gained over time through the interpenetration of polymer chains via reptation processes. Our experimental data indicate that the strength gained over time follows the power-law $(\sigma - \sigma_0) = t^\alpha$, with $\alpha = 0.18$ (Figure S6). The scaling exponent of 0.18 is relatively close to the value of 0.25 predicted by theory, suggesting that strength recovery in the self-healing silicone vitrimer might be governed by the interpenetration of polymer chains across the cut surfaces, which in turn enables the occurrence of the metathesis reactions between dioxaborolane moieties.

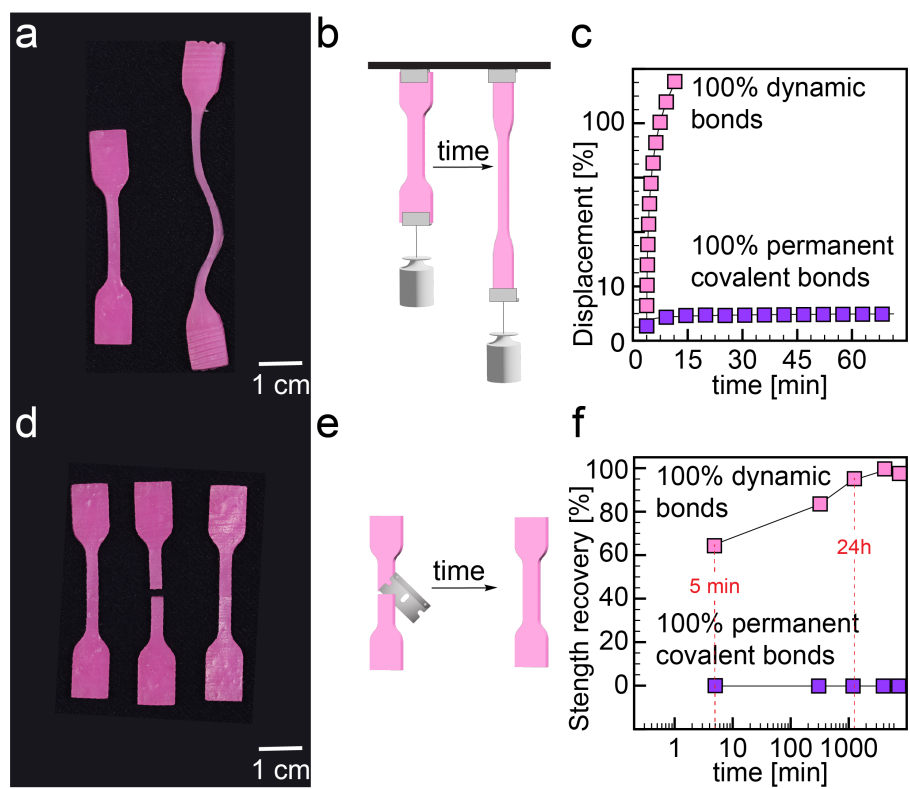

**Figure 3. Creep and self-healing of printed silicones with dynamic or permanent bonds**. (a,d) Photographs of the printed dogbone specimens used for creep (top) and self-healing tests (bottom). (b,e) Schematic representation of (b) creep and (e) self-healing experimental procedures. (c) Creep data expressed in terms of the relative displacement of the tensile-loaded sample as a function of time. (d) Evolution of the self-healing efficiency of samples that were intentionally cut and put back in contact for different durations before testing.

Silicone architectures that synergistically combine self-healing and creep-resistant properties can be obtained by multimaterial printing inks with silicone formulations containing dynamic or permanent covalent bonds. To illustrate this concept, we first printed simple bilayer architectures consisting of a self-healing layer containing dynamic covalent bonds and a creep-resistant layer with permanent covalent bonds (Figure 4a). In such a design, the presence of dynamic covalent bonds imbues one of the layers with self-healing capabilities, whereas the elastic recovery of the permanently crosslinked layer ensures close contact between fractured surfaces possibly formed in the healable layer. Despite the different geometries, this architecture closely resembles the design principles of the elastic, self-healing byssus threads of mussels. [11] Importantly, the use of dioxaborolanes as dynamic covalent bonds allows exchange reactions to take place at room temperature within the highly mobile silicone network, which is crucial for a fully autonomic self-healing behavior.

To quantify the synergistic properties of the architectured silicone, we perform self-healing and creep resistance experiments on printed bilayers (Figure 4a). The relative fraction of permanent and dynamic covalent bonds within the bilayer structure was varied by changing the relative thickness of the two layers. For the self-healing tests, a razor blade was used to introduce damage in the layer with dynamic bonds and the tensile strength of the bilayer was measured as a function of time. Since the dioxaborolane metathesis reaction responsible for the chemical exchange processes in the silicone can occur at room temperature (Figure 4b), network covalent bonds are expected to reform in the healable layer as soon as the two cut parts are put into contact by the underlying elastic layer (Figure 4c).

For a bilayer structure comprising 50% of silicone vitrimer and 50% of silicone elastomer, our experiments reveal a ~40% recovery of the initial strength in only 5 minutes after the healable layer has been cut. The self-healing efficiency of the architectured silicone rises up to ~90% within 24h after the intentional damage. Such response is not far from the self-healing performance shown by the printed single layers (Figure 4d). Notably, in the case of the bilayer no human intervention was necessary to put the cut surfaces into close contact. The high efficiency of the autonomic self-healing process is also reflected by the nearly unnoticeable scar left at the damaged site. When two dioxaborolane groups present on the two separate halves come into contact, they undergo a metathesis reaction that merges the chains of the damaged parts, thus creating new strong dynamic covalent bridges between the cut parts.

The creep resistance of the architectured silicone was assessed using samples with different relative thicknesses of the permanently and dynamically crosslinked layers (Figure 4e). The experimental results reveal that a relative thickness of 25% of the elastic layer is sufficient to limit the displacement of the stressed sample and thus prevent creep of the bilayer. Because of the stiffer nature of the elastic layer, the total deformation of the bilayer decreases with the relative fraction of permanent covalent bonds in the sample. In addition to creep resistance, the elastic layer also increases the tensile strength, mechanical stiffness and elastic recovery of the bilayer (Figure S7). Such a strengthening effect can be explained using a simple rule of mixture analysis (Figure S8).

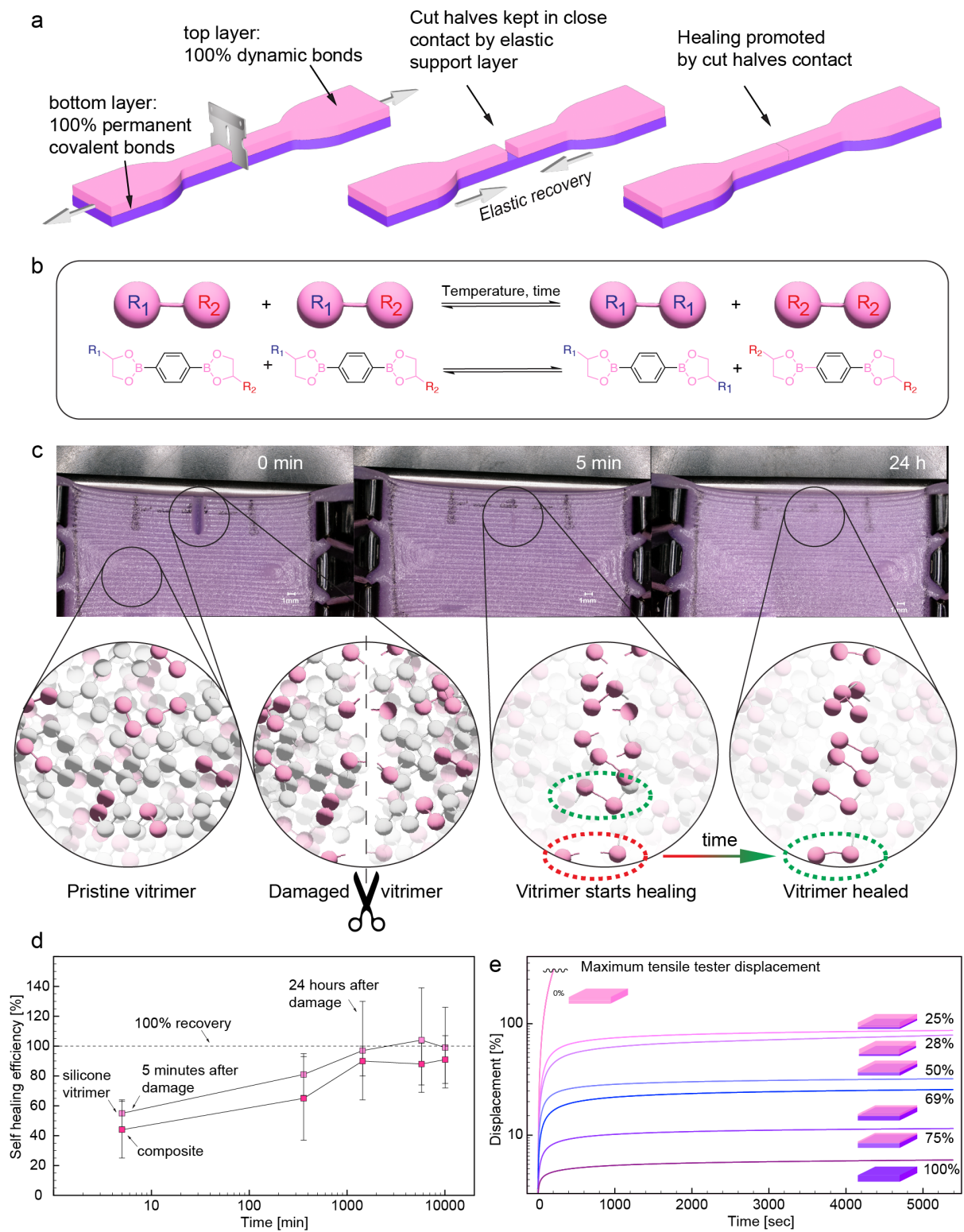

**Figure 4. Self-healing and creep-resistant architectured silicone.** (a) Cartoon depicting the synergetic combination of silicone vitrimer and silicone elastomer layers to form an architectured

silicone sample. (b) Exchange reactions between dioxaborolane groups of the silicone vitrimer network responsible for the self-healing effect. (c) Optical images of the dynamically crosslinked layer during self-healing of an intentionally introduced cut. The cartoons show the exchange reactions and topological changes that occur between dynamic crosslinks across the cut surfaces. (d) Evolution of the self-healing efficiency of the bilayer architectured silicone (50% vitrimer; 50% elastomer) compared to a reference sample containing only dynamic covalent bonds. (e) Displacement as a function of time obtained from creep tests on bilayers with different relative thickness of the silicone vitrimer and silicone elastomer layers.

The ability to resist creep deformation and self-heal at room temperature without external triggers makes the architectured silicones attractive for a broad range of applications in the form of resilient soft actuators, artificial robotic skins or conformable structures. To demonstrate this potential, we 3D printed a water-sealed soft tube with a trilayer silicone architecture that enables autonomic self-healing and recovery of its pumping functionality after damage. This geometry was chosen due to its high symmetry, which enables a more homogeneous inflation and deflation sequence when connected to a water stream in comparison to the aortic arch section. The trilayer architecture comprises a dense layer of silicone elastomer sandwiched between two external dense layers of silicone vitrimer.

The soft tube is produced through the sequential deposition of the vitrimer, the elastomer and the vitrimer inks using a 5-axis 3D printer (Figure 5a). A printing nozzle with diameter of 0.84 mm was used to reach a silicone thickness of 0.6 mm for each layer (Figure 5b). The printing path was designed based on a laser scan of the cylindrical substrate to be covered. After UV curing in $N_2$ atmosphere for 10 minutes, the multi-layered tube was sealed at its ends with 3D-printed customized components covered with raw resin. A final curing step was performed to obtain the water-sealed soft tube with architectured silicone walls.

In its pristine state, the soft tube can be inflated and deflated compliantly using water as driving fluid. To evaluate the ability of the soft tube to recover from damage and restore such inflating functionality, we intentionally cut the tube wall with a pair of scissors (Figure 5c). This led to a 2cm-long defect across the entire thickness of the wall. A few seconds after being damaged, the tube leaks water in a catastrophic way when reconnected to the water line. By turning off the water supply, the cut halves can be brought into close contact thanks to the elastic recovery of the silicone elastomer.

Self-healing of the tube walls occurs within less than 5 minutes through the dioxaborolane metathesis reaction within the silicone vitrimer network. Three minutes of healing is already enough to partially fix the damage and reduce water leakage. Letting the tube rest for a few more minutes leads to complete closure of the cut, enabling water to flow again through the tube. The initial scar left on the architectured wall is eventually blended into the tube's texture after many cycles of inflation and deflation. Notably, the room-temperature healing of the two silicone vitrimer layers is sufficient to restore the function of the vascular-like tube.

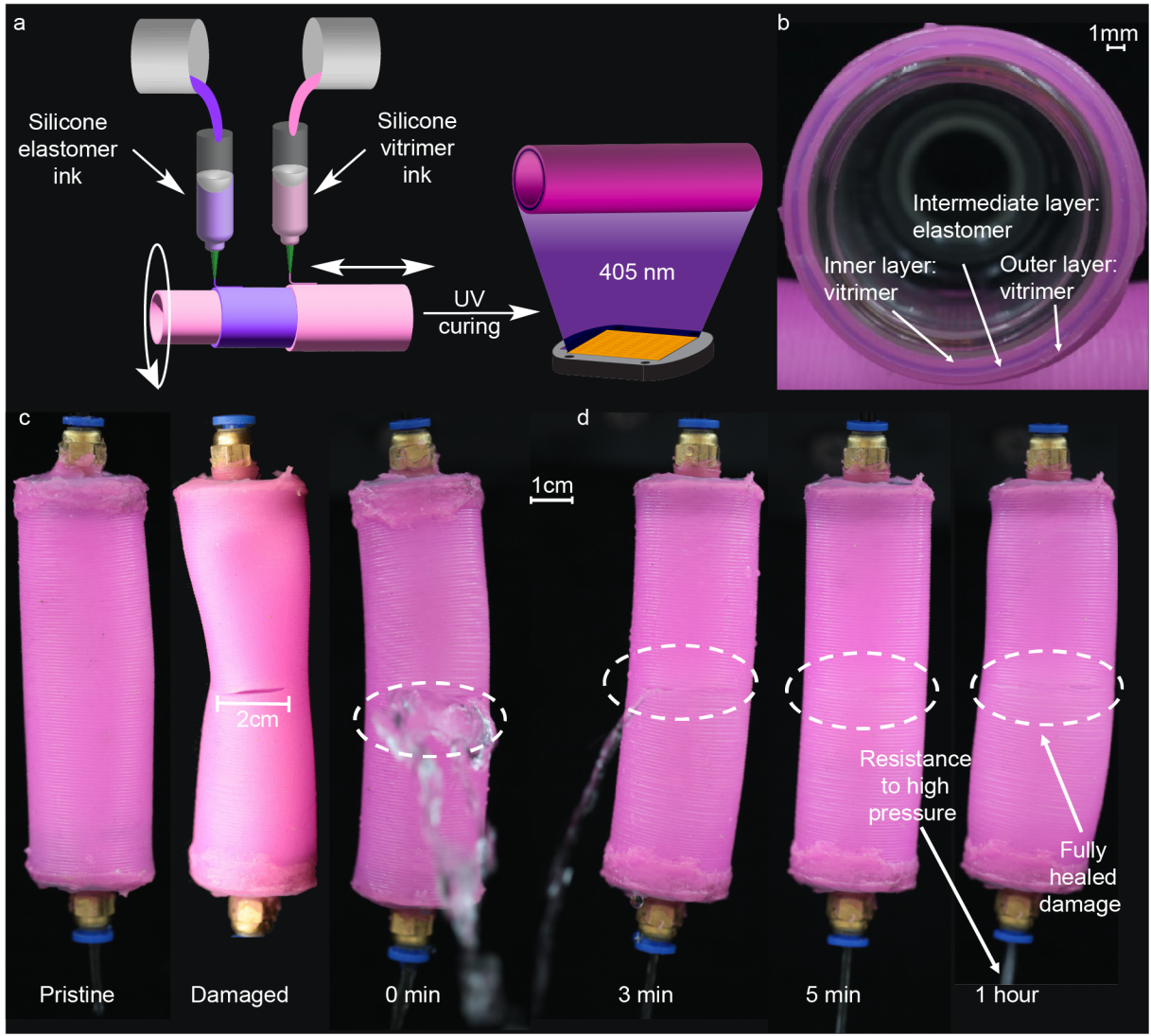

**Figure 5.** Vascular phantom silicone tubes with self-healing properties at room temperature. (a) Schematics of the multimaterial direct ink writing process used to create a soft tube with trilayered silicone architecture. (b) Cross-section of the layered tube, highlighting the elastic inner layer sandwiched between two external self-healing layers. (c) Photographs of the pristine (left) and damaged (right) silicone tubes. (d) Snapshots depicting the water leakage shortly after damage and the ability of the self-healing architectured walls to restore the tube's function in just 5 minutes.

**Conclusions**

Architectured silicones printed from vitrimer and elastomer siloxane-based inks can be designed to showcase creep resistance and autonomic self-healing, two properties that are often antagonistic in state-of-the-art materials. Creep resistance arises from the presence of permanent covalent bonds within the silicone network layer, whereas self-healing is achieved by a vitrimer layer comprising dynamic covalent bonds based on dioxaborolane chemistry. Self-healing at room temperature is possible due to the synergetic interplay between the permanent and dynamic covalent bonds present in different silicone layers. The elastic energy stored in the elastomer network is harnessed to establish contact between the fractured surfaces of the vitrimer layer, thus allowing for exchange reactions between the dioxaborolane groups to take place and self-healing to proceed without external triggers. Because of the high mobility of the silicon chains, 40-50% of the initial strength of the vitrimer is readily recovered upon contact of the damaged surfaces. Further healing and strength recovery is controlled by diffusion of silicone chains within the dynamic covalent network. The formation of dynamic and permanent covalent networks with sufficient strength and stiffness was possible by synthesizing thiol-based prepolymers of relatively high molecular weight using a pre-polymerization and chain-extension protocol. Mixing such polymers with a crosslinker, photoinitiator and fumed silica particles resulted in photo-curable inks that can be printed into three-dimensional objects with complex architectures using the direct ink writing technique. Using a multimaterial approach, we printed a vascular-mimicking surgical training aid that is able to autonomously recover its strength and function through the self-healing of architectured silicone walls. The ability to self-heal without human intervention combined with creep resistance makes such architectured silicones attractive materials for potential applications in soft robotics, surgical training and stretchable electronics.

## Acknowledgements

We thank the funding kindly provided by the Strategic Focus Area Advanced Manufacturing (SFA-AM) of the Swiss ETH domain. The authors are grateful to Prof. Dr. Athina Anastasaki for fruitful discussions, Dr. Marco Binelli and Alessandro Dutto for their support with rheology and Dr. Fabio Bargardi for his help with designing and manufacturing the UV curing system. We also thank Prof. Dr. Eric Dufresne and his group for providing access to their uniaxial tensile tester, Prof. Dr. Markus Niederberger and Dr. Marco d'Elia for providing access to their GPC setup.

## Materials and Methods

**Materials**

Vinyl-terminated polydimethylsiloxane (DMS-V22, 200 cSt) and (7.0-8.0% vinylmethylsiloxane)-dimethylsiloxane copolymer, trimethylsiloxy terminated (VDT-731, 800-1200 cSt) were purchased from Gelest, Inc. (USA). Thioglycerol ($\geq$ 97%), tetrahydrofuran (THF, $\geq$ 99.9%), toluene ($\geq$ 99.7%), chloroform-d (99.8%) and molecular sieves (zeolite, 4Å) were acquired from Sigma-Aldrich (USA). 1,6-Hexanedithiol (97%) was purchased from AlfaAesar (USA), whereas (1,4-Phenylenediboronic acid (98%) was obtained from Combi-Blocks, Inc. (USA). 2,2-Dimethoxy-2-phenylacetophenone (DMPA, 99%) was purchased from Acros Organics (USA). SilcPig$^{TM}$ silicone color pigments were acquired from Smooth-On Inc. (USA), while fumed silica (Aerosil R8200) was purchased from Evonik AG (Germany).

**Synthesis methods**

Synthesis of monomer containing dioxaborolane functional groups

The monomer (1,4-phenylenebis(1,3,2-dioxaborolane-2,4-diyl))dimethanethiol was prepared via condensation of 1,4-phenylenediboronic acid with thioglycerol (Figure S9). First, 16.00g (96.48 mmol, 2eq) of 1,4-phenylenediboronic acid were added to a 2000ml 2-neck round bottom flask, followed by the slow addition of 1440ml of toluene under stirring until complete dissolution. The flask containing this milky solution was then connected a Soxhlet extractor equipped with molecular sieves and slowly warmed up to 60 °C. After 10 minutes, 20.88g (193.04 mmol, 4eq) of thioglycerol were added dropwise while keeping the temperature at 110°C. The solution turned completely clear in a few minutes, and it was left under vigorous magnetic stirring at reflux temperature. After 18h, the reaction solution was filtered and the final product was concentrated by rotary evaporation without further purification to obtain a coarse white powder (28.96g, yield ~82%. $^1$H NMR (300MHz, CDCl$_3$): δ 7.82 p.p.m (s, 4H), 4.74 (m, 2H), 4.49 (m, 2H), 4.17 (m, 2H), 2.81 (m, 4H), 1.50 (dd, J=11, 7.4 Hz, 2H) (Figure S10)).

Synthesis of silicone vitrimer prepolymer

In a 250ml single-neck round-bottom flask, 6.5g (20.9 mmol, 3eq) of (1,4-phenylenebis(1,3,2-dioxaborolane-2,4-diyl))dimethanethiol were slowly dissolved with 120mg (0.46 mmol) DMPA in 150ml of THF. After complete dissolution of the solids, 66.3g (7.0 mmol, 1eq) of vinyl-terminated polydimethylsiloxane were slowly added under magnetic stirring. The solution was then purged with argon for 3 minutes. The sealed flask was then placed inside a home-built, air-tight UV curing chamber under N$_2$ atmosphere. The solution was kept under magnetic stirring and simultaneously irradiated for 10 minutes using a mercury lamp (Omnicure S2000 UV lamp 200W, 320-500nm) from the side and using a home-built LED system (50W, 405nm) purchased from AliExpress (China) from the bottom. For both UV sources, the measured intensity was >200mW/cm$^2$. The solution was then concentrated by rotary evaporation to obtain a viscous, milky solution.

Synthesis of silicone elastomer prepolymer

In a 250ml single-neck round-bottom flask, 2.3g (15.3mmol, 2eq) of 1,6-hexanedithiol were slowly dissolved with 130mg (0.49 mmol) DMPA in 160ml THF. After both chemicals were dissolved, 73.8g (7.7mmol, 1eq) of vinyl-terminated polydimethylsiloxane were slowly added under stirring. The solution was then purged with argon for 3 minutes. The sealed flask was then placed inside the UV curing chamber and, under magnetic stirring, irradiated simultaneously with the mercury lamp from the side and with the LED system from the bottom for 10 minutes. In both cases, the UV intensity was measured to be >200mW/cm$^2$. The solution was then concentrated by rotary evaporation to obtain a viscous, transparent solution.

**Ink formulations for direct ink writing**

Silicone vitrimer

In a 300ml Thinky mixer cup (Thinky, USA), 69.23g of silicone vitrimer prepolymer were added to 46.15g of trimethylsiloxy-terminated (7.0-8.0% vinylmethylsiloxane)-dimethylsiloxane copolymer, 225mg DMPA (0.15 wt%) and 300mg (0.2wt%) SilcPig (color: electric pink). The fraction of vinylmethylsiloxane in the (vinylmethylsiloxane)-dimethylsiloxane copolymer was 7-8%. The mass ratio of 2:3 between the silicone vitrimer prepolymer and the vinyl-siloxane copolymer was optimized empirically to maximize the mechanical performance of the silicone vitrimer. After manual mixing for 1 minute, 34.61g (30 parts per hundred of rubber, phr) of silica nanoparticles (Aerosil R8200) were slowly incorporated into the resin formulation and mixed for 3 minutes at 2200rpm in a planetary centrifugal mixer (Thinky Mixer ARE-250) to obtain approximately 150g of 3D-printable silicone-based resin.

Silicone elastomer

In a 300ml Thinky mixer cup, 70.31g of silicone elastomer prepolymer were added to 23.43g of trimethylsiloxy-terminated (7.0-8.0% vinylmethylsiloxane)-dimethylsiloxane copolymer, 225mg DMPA (0.15 wt%) and 300mg (0.2 wt%) SilcPig (color: electric violet). The fraction of vinylmethylsiloxane in the (vinylmethylsiloxane)-dimethylsiloxane copolymer was 7-8%. The mass ratio of 2:3 between the silicone vitrimer prepolymer and the vinyl-siloxane copolymer was optimized empirically to maximize the mechanical performance of the silicone elastomer. After manual mixing for 1 minute, 56.25g (60phr) of silica nanoparticles (Aerosil R8200) were slowly incorporated into the resin formulation and mixed for 3 minutes at 2200rpm in a Thinky mixer to obtain approximately 150g of 3D-printable silicone-based resin.

**3D-printing and curing**

To 3D-print complex structures via direct ink writing we used 30 cm$^3$ syringe barrels (model ADV830BA) equipped with syringe pistons (model ADV830WW) and with a 0.84mm nozzle (model TT18-rigid). All 3D-printing consumables were purchased from Adhesive Dispensing Ltd., UK. The syringe barrels were filled by hand up to 75% of their full capacity and then centrifuged for 3 minutes before being connected on different Direct-Ink-Writing 3D printers.

We used an- extrusion-based printer (3D-discovery, RegenHU SA, Switzerland) to print 1.8mm-thick films using distinct ink formulations. Dogbone samples were extracted from these films and used in the quasi-static, creep, elastic recovery and self-healing tests. The inks were printed with an applied pressure of approximately 3.5bar at a feed rate of 12mm/s.

A custom-built 5-axis 3D printer was utilized to fabricate the self-healing tubular actuator on a cylindric substrate and to print the aortic phantom model on a CT scan derived mandrel. The substrate surface was first scanned with a laser measurement device (Keyence LK-G5000) and the printing path was generated using in-house software. The inks were printed with an eco-PEN300 extruder, a flow rate of 210ml/min and a machine feed rate of 280mm/min.

All printed parts were then transferred inside the air-tight UV chamber and placed approximately 5mm away from the UV source. After flushing for 3 minutes with nitrogen, the UV lamp (Omnicure S2000) was turned on and the parts were illuminated for 10 minutes under a slightly positive $N_2$ pressure with a light intensity of approximately 200mW/cm$^2$.

**Ink rheology**

All rheological tests were carried out at 25°C on a stress-controlled compact rheometer (Anton Paar MCR 302 rheometer) using a sandblasted parallel plate geometry (PP25) with a 1 mm gap. The apparent viscosity of the resins was measured under steady-shear conditions by increasing the applied shear rate from $10^{-3}$ to 1 s$^{-1}$, followed by a decrease in shear rate from 1 to $10^{-3}$ s$^{-1}$. Only the data obtained under decreasing shear rates is shown in Figure S3a.

Elastic recovery tests were performed by alternating between oscillatory and steady-shear measurements. The oscillatory measurements were conducted at a maximum amplitude strain of 1% and frequency of 10 rad/s, whereas the steady-shear tests were performed at a shear rate of 0.1 s$^{-1}$. While these conditions do not take into account the extensional flow that occurs during extrusion, they were used as a proxy for the forces applied to the inks during the 3D printing process.

To evaluate the viscoelastic behavior of the inks, we carried out measurements in stress-control mode starting at a minimum stress of 1 Pa and using an integration of 10s per point.

**Mechanical characterization of 3D-printed parts**

All mechanical tests were carried out at room temperature on dogbone-shaped samples with sizes recommended by the ISO 37 (type 4) standard. Measurements were performed in a uniaxial tensile tester (Shimadzu AGS-X) using a 100N load cell for quasi-static, elastic recovery and self-healing tests. Creep tests were conducted in a uniaxial tensile tester (TA.XTPlus, Stable Micro Systems) using a 5N load cell. Quasi-static and self-healing tests were performed by stretching the specimen at 80mm/min until rupture.

For the self-healing tests on the silicone vitrimer, printed films were cut into half with a razor blade, immediately brought back into contact and left unloaded for precise time intervals before the mechanical measurements. The time intervals varied broadly from 5minutes to 30 days. The strength of the samples after this resting period was measured and compared to that of the pristine sample.

The same general protocol was adopted to test the self-healing efficiency of the architectured silicones, except for the procedure used to create the damaged site. For the architectured samples, only the silicone vitrimer layer was cut with the razor blade, thus leaving the silicone elastic layer untouched. The modulus at 100% elongation (M100) was determined between strain values of 99 and 101%. Reported values for strength, strain at break and M100 are averaged over at least three samples (Figure S8).

Elastic recovery tests were performed by stretching the specimen at a speed of 80mm/min until 140% elongation and slowly unloading the samples with a speed of 1mm/min until a force as low as 0.02N was measured. Elastic recovery was then calculated as the percentage ratio between the recovered deformation after unloading ($\varepsilon_{max} - \varepsilon_{0.02}$) and the maximum deformation imposed ($\varepsilon_{max}$).

Creep was measured by first applying an initial stretch up to 0.2MPa at a rate of 80mm/min, followed by a static load of 0.2MPa for 90 minutes.

**Nuclear Magnetic Resonance and Gel Permeation Chromatography**

$^1$H NMR spectra were recorded on a Bruker AV 300 MHz (Billerica, MA, USA) spectrometer in CDCl$_3$. Gel permeation chromatography (GPC) analysis were carried out using a Viscotek GPC system (Malvern, Worcs, UK) equipped with a pump and degasser (GPCmax, VE2001, 4.0 ml/min flow rate), a detector module (Viscotek 302 TDA) and three columns (2x PLGel Mix-C and 1x ViscoGEL GMHHRN 18055, dimensions 7.5 x 300 mm for each column) using THF as eluent.

# Supporting Information

**3D Printed Architectured Silicones with Autonomic Self-healing and Creep-resistant Behavior**

Stefano Menasce, Rafael Libanori, Fergal Coulter, André R. Studart

Complex Materials, Department of Materials, ETH Zürich, 8093 Zürich, Switzerland

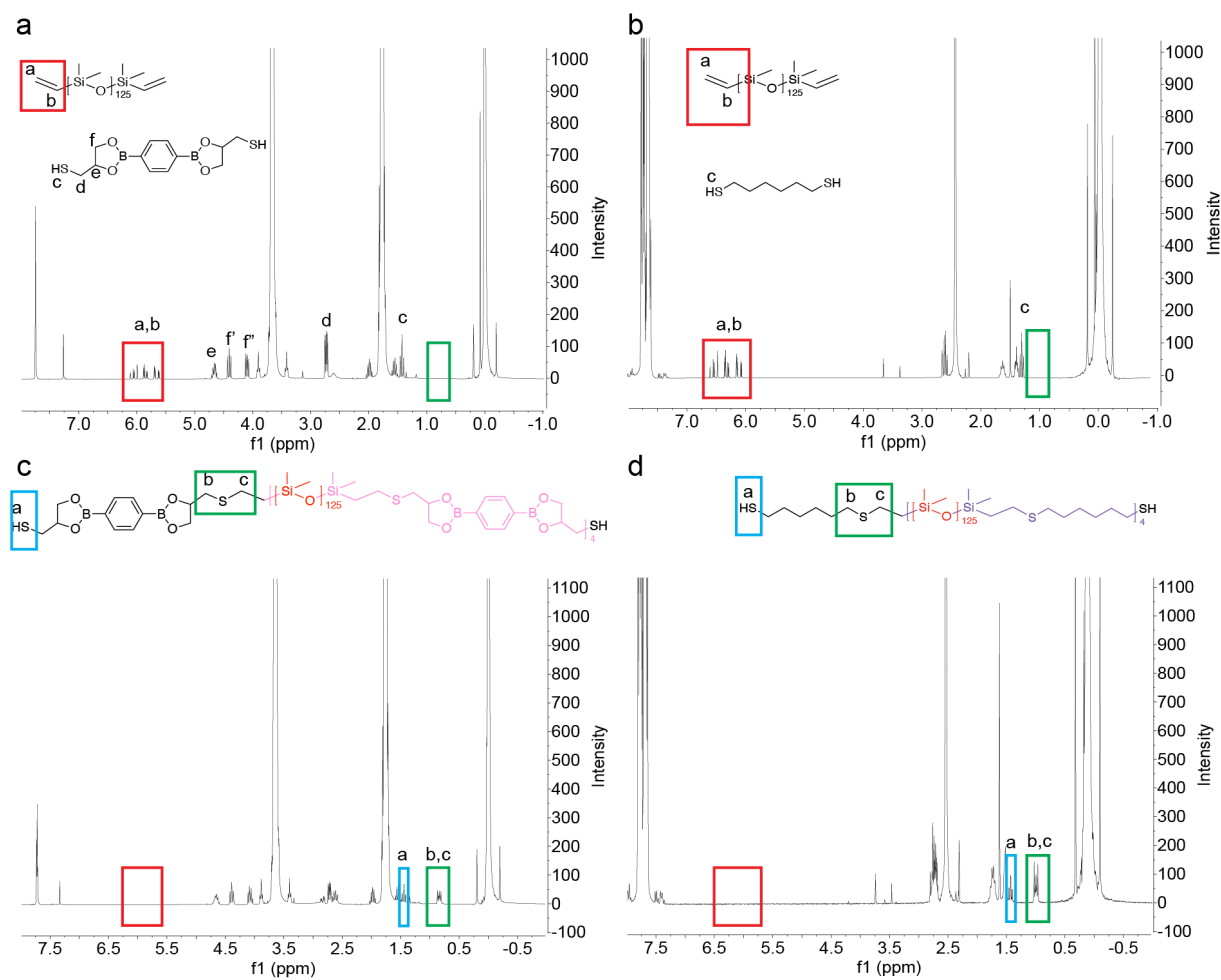

**Figure S1.** $^1$H NMR spectra of the thiol-based monomers and the silicone prepolymers obtained after the chain extension reactions. Spectra of the monomers with (a) dynamic covalent bonds and (b) permanent covalent bonds. Spectra of (c) the silicone vitrimer prepolymer and (d) the silicone elastomer prepolymer. The highlighted regions indicate the disappearance of vinyl peaks (red) and the appearance of sulfide peaks (green) after the pre-polymerization reaction.

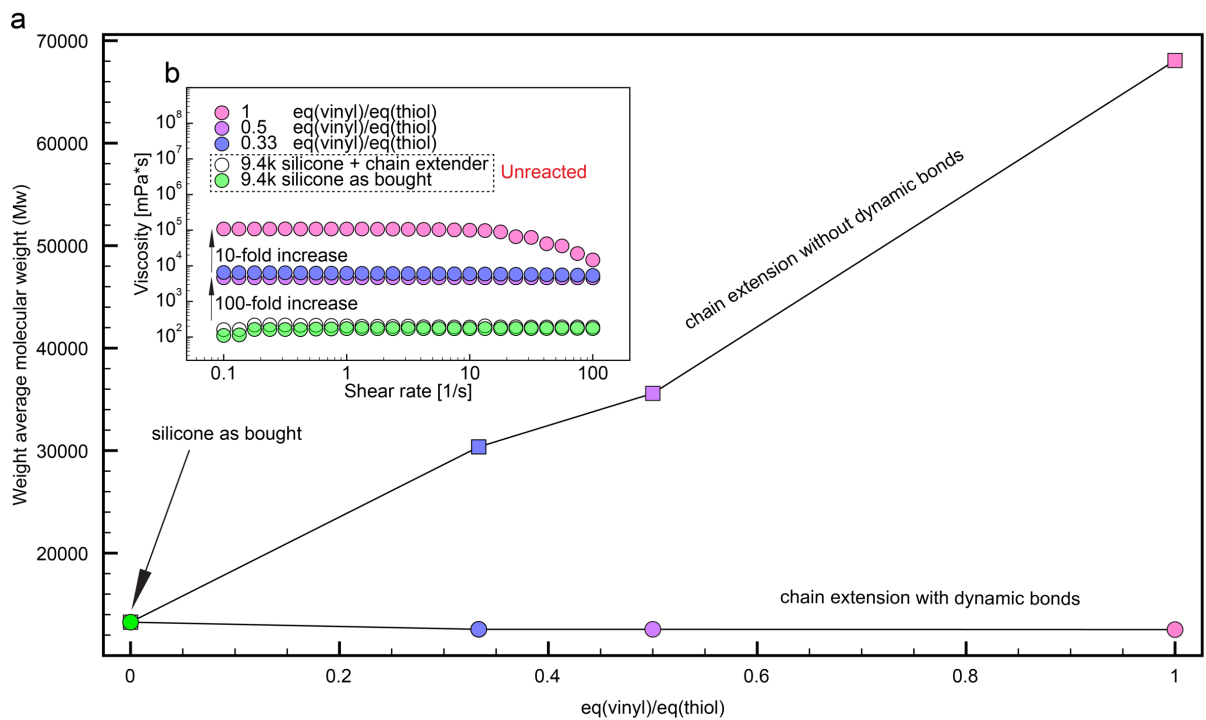

**Figure S2**. a) Molecular weight ($M_w$) of silicone polymers before (green spot) and after (colored shapes) chain extension, as a function of the equivalents ratio of vinyl and thiol group in solution. Squares represent the silicone polymers chain-extended using 1,6-hexanedithiol, while spheres represent the silicone polymers chain-extended using 1,4-phenylenebis(1,3,2-dioxaborolane-2,4-diyl))dimethanethiol. b) Apparent viscosity of silicone chain-extended using 1,4-phenylenebis(1,3,2-dioxaborolane-2,4-diyl))dimethanethiol with different ratios of vinyl and thiol equivalents. The rheological data is compared to those obtained for as-bought silicone prepolymer or after mixing it with chain extender. While no significant difference is found between compositions with ratios of 0.33 and 0.5, an approximately 10-fold increase is observed when the stoichiometric ratio is achieved. In terms of molecular weight, the use of 1,6-hexanedithiol as chain extender increases the $M_w$ value of the silicone polymer in proportion to the eq(vinyl)/eq(thiol) ratio up to the stoichiometric value. In contrast, an increase in molecular weight is not observed when using 1,4-phenylenebis(1,3,2-dioxaborolane-2,4-diyl))dimethanethiol as chain extender. The presence of acidic hydroxyl groups on the silica contained in the GPC column might induce fragmentation of the hydrolysable boronic esters groups of the dynamic silicone and thus explain the observed constant molecular weight of chain-extended dynamic silicone. The formation of the disulfide groups in the first place was confirmed by NMR (Figure S1d) and by the significant increase of the polymer's viscosity when approaching the stoichiometric equivalent ratio (Figure S2a).

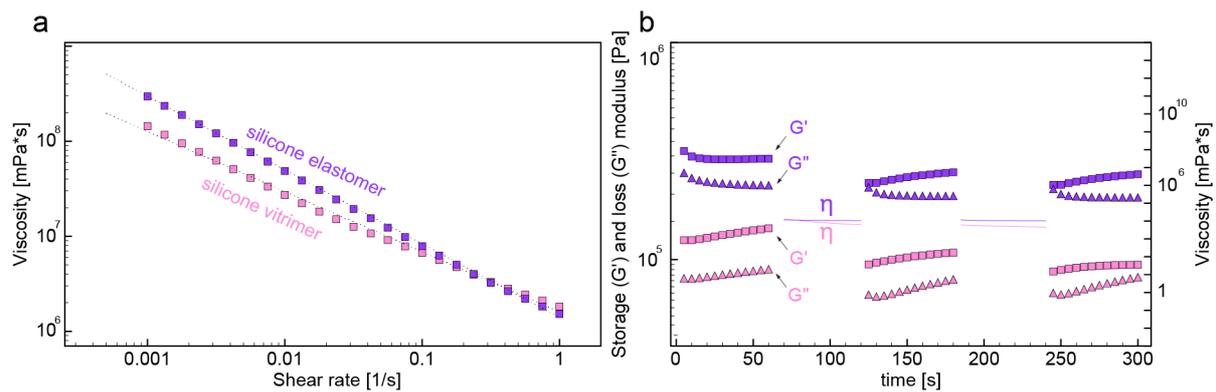

**Figure S3.** Rheological behavior of the silicone inks containing prepolymers with dynamic or permanent covalent bonds. (a) Apparent viscosity as a function of shear rate obtained from steady-shear measurements on silicone resins before curing. (b) Elastic recovery tests designed to simulate the extrusion of the resin through the nozzle during printing. Oscillatory shear measurements are performed for 60s, before and after 60s of steady-shear flow, to quantify the initial and recovered viscoelastic properties of the resins.

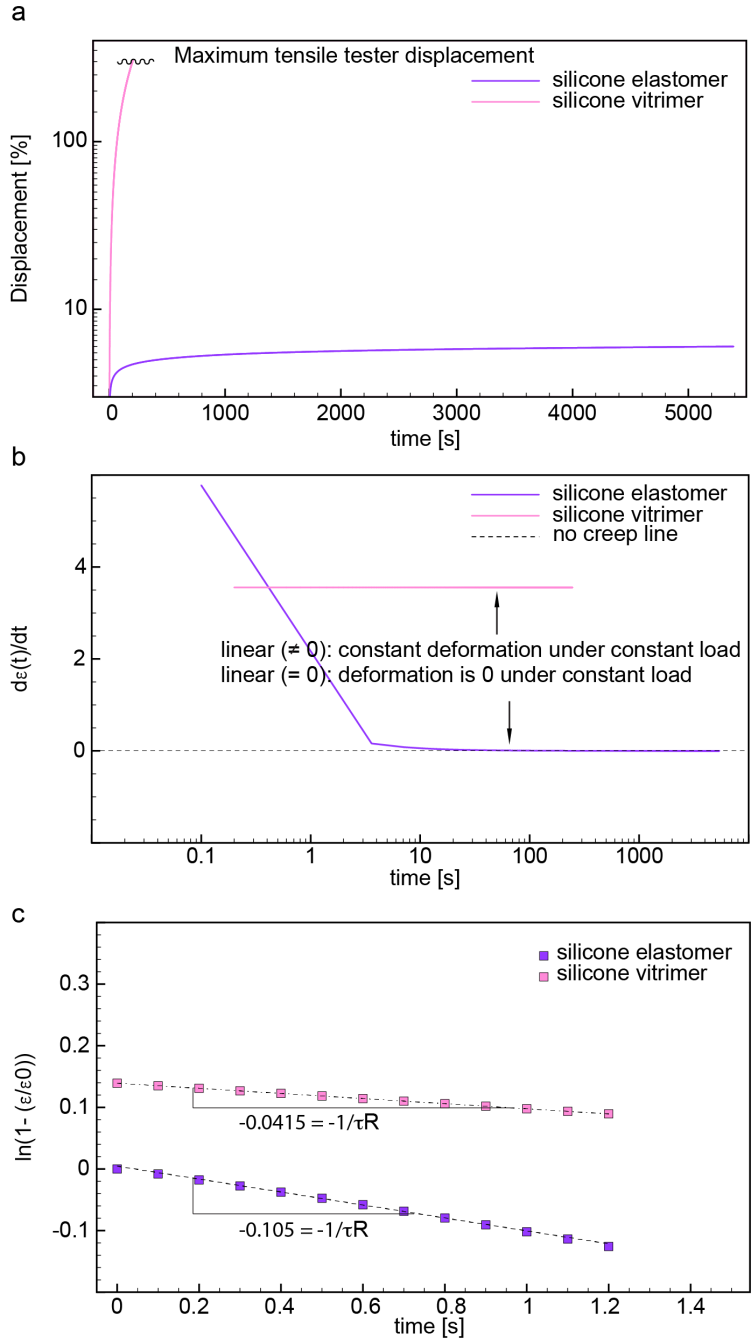

**Figure S4.** Creep behavior of the silicone vitrimer and the silicone elastomer. (a) Strain of the silicone samples as a function of time for a constant applied tensile stress ($\sigma_0$) of 0.2 MPa. (b) Creep rate calculated by taking the slope of the raw data shown in (a). (c) Plot of $ln(1-(\varepsilon/\varepsilon_0))$ versus time ($t$) used to determine the retardation time $\tau_R$ for the silicone elastomer and the silicone vitrimer upon mechanical loading of the samples at the beginning of the creep test. In this analysis, the strain of the material under constant stress ($\varepsilon(t)$) is assumed to follow the Kelvin-Voigt model: $\varepsilon(t) = (\sigma_0/E)(1-e^{-t/\tau_R})$, where $(\sigma_0/E) = \varepsilon_0$ and $E$ is the elastic modulus. Based on this assumption, the retardation time can be determined from the slope of the plot of $ln(1-(\varepsilon/\varepsilon_0))$ versus time, which is equal to $-1/\tau_R$. Linear fittings to the experimental data obtained at the early stage of the deformation process leads to retardation times of approximately 9.5s and 24.1s for the silicone elastomer and the silicone vitrimer, respectively. The 5-fold higher relaxation time estimated for the silicone vitrimer reflects the more dissipative character of this dynamic covalent network compared to the permanent counterpart.

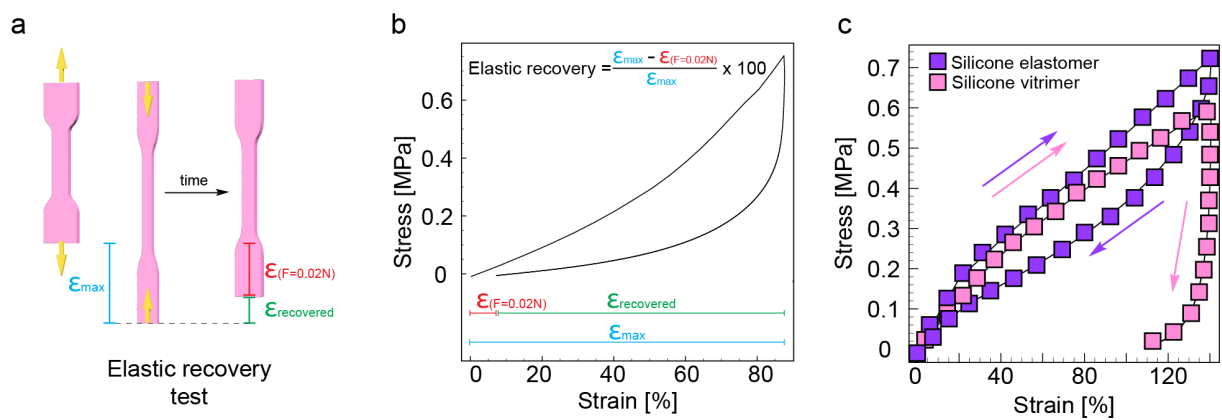

**Figure S5.** Elastic recovery measurements of silicones with dynamic or permanent covalent bonds. (a,b) Schematics depicting the parameters measured during the tests and the equation used to calculate the elastic recovery of the silicone sample. Each sample is elongated to 140% of its initial length and slowly unloaded until a force of 0.02N is detected. (c) Stress-strain curves obtained from elastic recovery tests performed on silicones with dynamic or permanent covalent bonds.

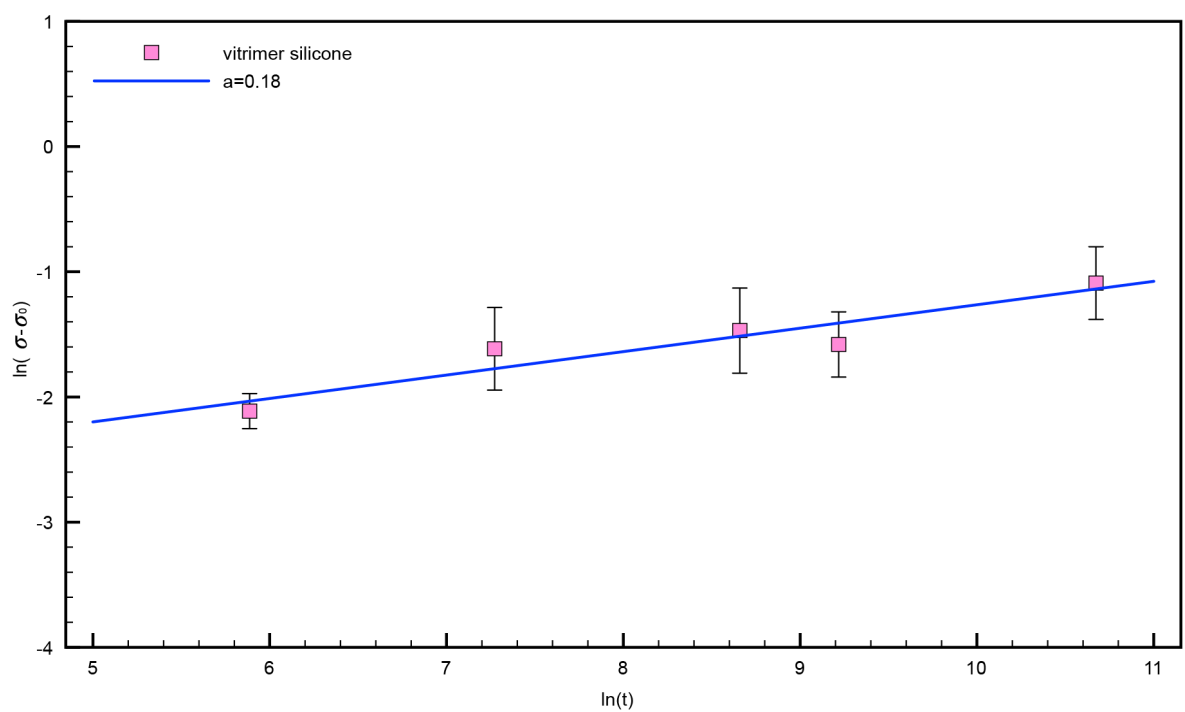

**Figure S6.** Analysis of the strength recovered during self-healing of the silicone vitrimer. The contribution of polymer chain diffusion to the recovered strength $(\sigma - \sigma_0)$ is plotted against the elapsed time $(t)$ in a double logarithmic scale. Linear fitting of the data allows us to determine the exponent $\alpha$ of the scaling relation: $(\sigma - \sigma_0) \sim t^\alpha$. The initial strength, $\sigma_0$, is taken as the strength value measured 5 mins after the samples had been put in contact. The exponent of 0.18 obtained from such fitting is relatively close to the theoretical $\alpha$ value of 0.25 predicted for strength recovery dominated by chain reputation via double random walk processes. [22]

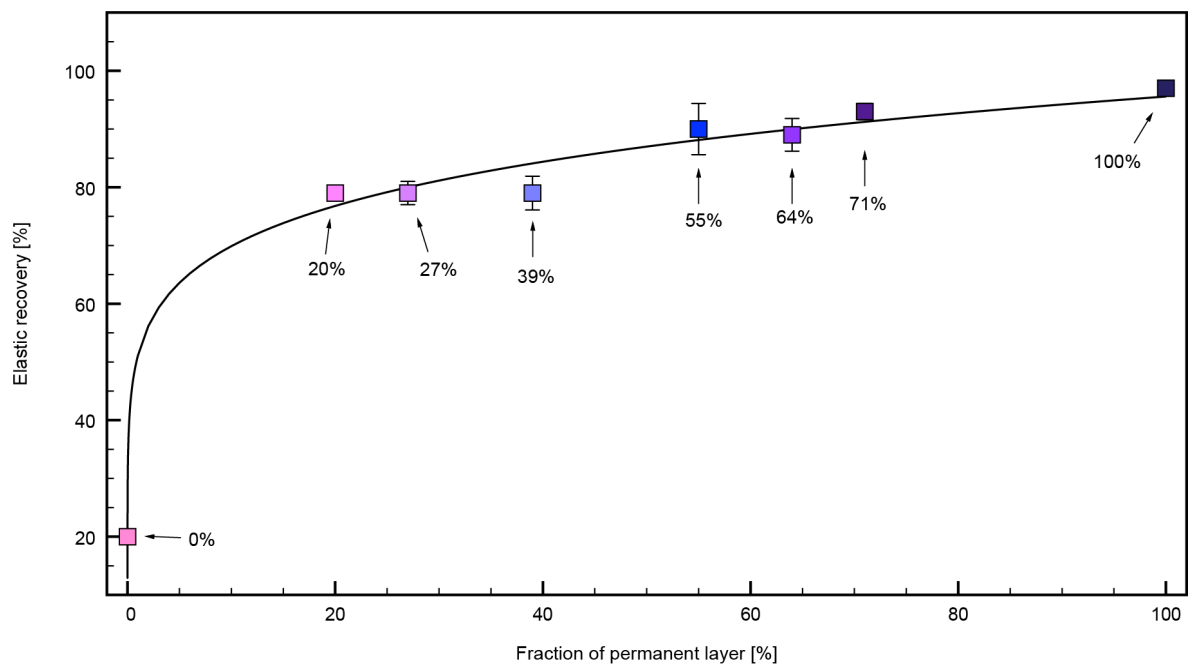

**Figure S7.** Elastic recovery of architectured silicones with different relative thickness of their constitutive layers.

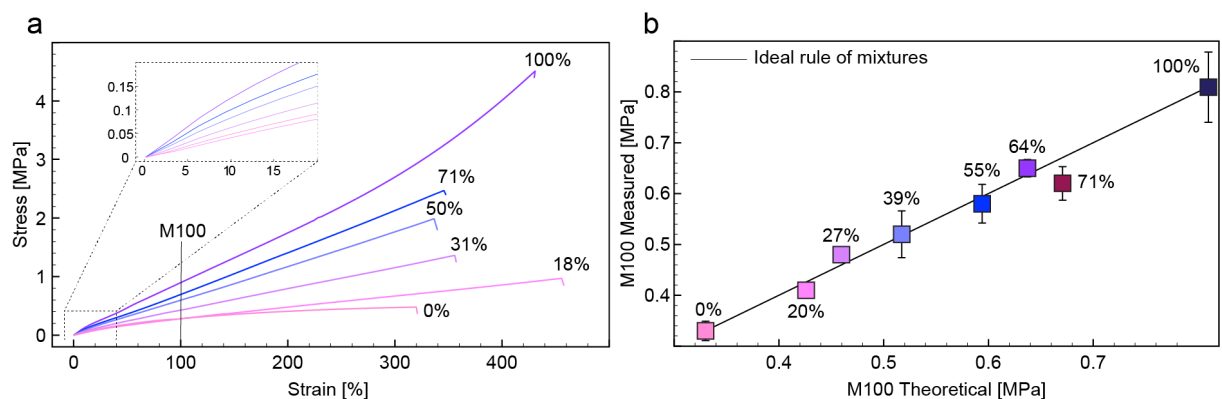

**Figure S8.** Mechanical properties of bilayer architectured silicones. (a) Stress-strain curves obtained for bilayer samples with different relative thicknesses of silicone vitrimer and silicone elastomer. (b) Tangent modulus at 100% strain (M100) obtained from the stress-strain curves compared with the values expected from a simple rule of mixture.

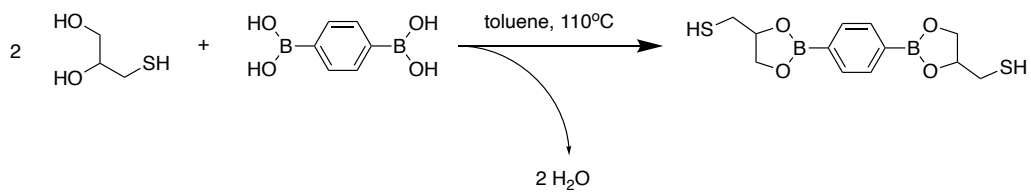

**Figure S9.** Schematics of the synthesis reaction leading to (1,4-phenylenebis(1,3,2-dioxaborolane-2,4-diyl))dimethanethiol.

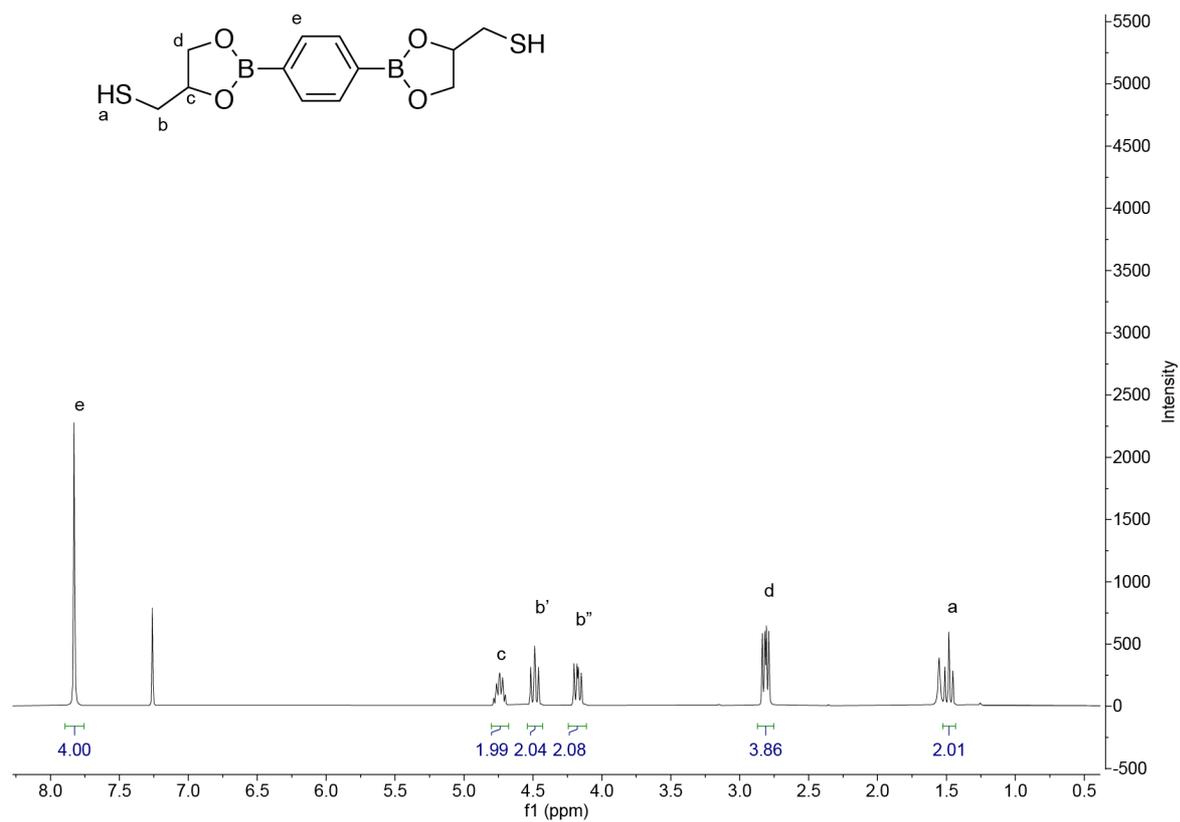

**Figure S10.** $^1$H NMR spectrum of (1,4-phenylenebis(1,3,2-dioxaborolane-2,4-diyl))dimethanethiol).